\documentclass[12pt,a4paper,oneside,onecolumn]{article}

\usepackage[a4paper]{geometry}

\usepackage{lipsum}
\usepackage{times}
\usepackage{amsmath}
\usepackage{amssymb}
\usepackage{float}
\usepackage[english]{babel}
\usepackage[utf8]{inputenc}
\usepackage[colorinlistoftodos]{todonotes} 
\usepackage[labelfont=bf]{caption}

\usepackage{csquotes}
\usepackage[style=authoryear,backend=biber]{biblatex}
\emergencystretch=\maxdimen
\usepackage{tocloft}
\usepackage{booktabs}
\usepackage{hyperref}
\usepackage[titletoc]{appendix}
\usepackage{tcolorbox}

\usepackage{graphicx}
\usepackage{lscape}
\usepackage{afterpage}

\title{Copula-Based Trading of Cointegrated Cryptocurrency Pairs}
\author{%
  \textsc{Masood Tadi}
  \thanks{Corresponding author} 
\\[1ex] 
\normalsize Faculty of Mathematics and Physics, Charles University \\
\normalsize Prague, Czech Republic\\ 
 \normalsize \href{mailto:tadim@karlin.mff.cuni.cz}{tadim@karlin.mff.cuni.cz} 
\and 
 \textsc{Jiří Witzany}
 \\[1ex] 
 \normalsize Faculty of Finance and Accounting, Prague University of Economics and Business\\
 \normalsize Prague, Czech Republic\\ 
 \normalsize \href{mailto:jiri.witzany@vse.cz}{jiri.witzany@vse.cz}
}
 \date{}

\usepackage{bbm} 
\usepackage{threeparttable}

\begin{document}
    \maketitle
	\setcounter{secnumdepth}{5}	
	\setcounter{tocdepth}{4}
	\pagenumbering{roman}
    \pagenumbering{arabic}
    \section*{Abstract}
    This research introduces a novel pairs trading strategy based on copulas for cointegrated pairs of cryptocurrencies. To identify the most suitable pairs, the study employs linear and non-linear cointegration tests along with a correlation coefficient measure and fits different copula families to generate trading signals formulated from a reference asset for analyzing the mispricing index. The strategy's performance is then evaluated by conducting back-testing for various triggers of opening positions, assessing its returns and risks. The findings indicate that the proposed method outperforms buy-and-hold trading strategies in terms of both profitability and risk-adjusted returns.
    
    \textbf{Keywords:} Statistical arbitrage, Pairs trading, Cointegration, Copulas, Cryptocurrency market.\\
    \textbf{Acknowledgements:} This paper has been prepared under financial support of a grant GAČR 22-19617S “Modeling the structure and dynamics of energy, commodity and alternative asset prices”, which the authors gratefully acknowledge. 
    
    \section{Introduction}\label{ch 1}
        Pairs trading is a well-known algorithmic trading strategy that capitalizes on temporary abnormal relationships among two or multiple assets whose historical prices tend to move together. When this relationship begins to exhibit abnormal behavior, it triggers the opening of trading positions. These positions are closed as soon as the pairs return to their normal behavior \parencite{vidyamurthy2004pairs}. In the decentralized cryptocurrency market, pairs trading can be profitable, offering two potential arbitrage opportunities: exchange-to-exchange arbitrage and statistical arbitrage. However, implementing a statistical arbitrage strategy based on exchange-to-exchange arbitrage can be risky and pose numerous challenges. In contrast, statistical arbitrage opportunities present similar profit potentials with lower risk \parencite{pritchard2018digital}.
    
        As stated by \textcite{krauss2017statistical}, pairs trading is characterized by a formation and a trading period. During the formation period, the objective is to identify pairs of assets that exhibit similar price movements. This is commonly achieved through co-movement criteria, which can be measured using various methods such as distance metrics (distance approach), for instance, the minimum sum of squared distances of normalized asset prices, or statistical relationships like cointegration rules (cointegration approach). Furthermore, the parameters of the trading period are estimated during the formation period. During the trading timeframe, irregularities in pairs' price movement are aimed to benefit from statistical arbitrage opportunities and create signals to open long/short positions. Advanced strategies may utilize a range of mathematical tools, such as stochastic processes, stochastic control techniques, copulas, and machine learning methods, to enhance the effectiveness of their outcomes.

         This paper aims to implement a copula-based pairs trading strategy on cointegrated cryptocurrency pairs. First, we start with a literature review on existing pairs trading strategies. We also provide an overview of the cryptocurrency exchange and data sources used in this study. The theoretical framework of the strategy includes linear and non-linear cointegration tests to identify cointegrated pairs and the copula concept to model their dependence structure. We discuss different families of copulas and the methods for copula estimation. We then outline the methodology for implementing the copula-based trading strategy, including ranking, selection of the assets, and the generation of the trading signals. Next, we present the empirical results of back-testing the strategy to a dataset of historical cryptocurrency prices. We analyze the performance of the strategy using various metrics, including profitability, risk-adjusted return, and maximum drawdown. In addition, we perform a comparative analysis of our results with the buy-and-hold strategy.
    \section{Literature Review}\label{ch 2}     
        There is extensive literature on pair trading strategies that use various concepts, such as the distance approach, cointegration analysis, or the concept of copulas. The distance approach involves calculating the historical price spread or price difference between two related assets and monitoring this spread over time. The spread is usually calculated as the difference between the prices of the two assets, either as a raw spread or as a normalized spread, such as the z-score or the percent difference. As an example, \textcite{gatev2006pairs} defined the normalized spread, $S^{ij}_{t}$, between assets $i$ and $j$ at time $t$ by 
        \begin{equation}
            S^{ij}_{t} = {P^i_t}^* - {P^j_t}^*\\
        \end{equation}
        \begin{equation}
        {P^i_t}^* = \frac{P^{i}_{t}}{P^{i}_{0}} =\prod_{\tau=1}^{t} \frac{P^{i}_{\tau}}{P^{i}_{\tau-1}} =\prod_{\tau=1}^{t}\left(1+r^{i}_{\tau}\right)= cr^{i}_{t} ,
        \end{equation}        
        where $P^{i}_{t}$ is the price of the asset $i$ at time $t$, $r^{i}_{t}$ is the t-period's return on asset $i$ at time $t$, and $cr^{i}_{t}$ is the cumulative total return of asset $i$ up to time $t$ ($cr^{i}_{0}=1 $). The calculation of the spread sum of squared distance is performed using the following equation:
        \begin{equation}
        SSD_{i,j}= \sum_{t=1}^{T}\left({S^{ij}_{t}}\right)^2=\sum_{t=1}^{T}(cr_{i,t}-cr_{j,t})^2.
        \end{equation}
        Pairs are chosen for the trading period based on their ascending order of $SSD_{i,j}$ during the formation stage. The initial pairs at the top of the list are selected, and basic non-parametric threshold rules are employed to generate trading rules. An alternative method was used by \textcite{chen2019empirical}, who identified the most suitable pairs for the trading period using Pearson correlation of assets' returns instead of finding the minimum sum of the squared distance. \textcite{krauss2017statistical} found that to maximize the profit of the distance approach strategy, the spread of each selected pair needs to have high volatility, indicating the potential divergence of the two assets. The pair's spread should also have a mean-reverting property. This approach's key advantage is its simplicity and transparency, making it suitable for large-scale empirical applications. 
        
        \textcite{huck2015pairs} found that the cointegration approach outperformed the distance approach in selecting effective pairs. The cointegration approach aims to identify a long-term equilibrium between non-stationary time series (e.g., asset prices) that move together. This equilibrium can be linear or nonlinear. \textcite{engle1987co} introduced the first cointegration test, which is based on linear regression and the unit-root test of residuals in the equilibrium. Typically, the augmented Dickey-Fuller test is used for the unit-root test. Other improvements of the Engle-Granger cointegration test were introduced by \textcite{phillips1990asymptotic} and \textcite{johansen1991estimation}. Highly volatile markets, such as cryptocurrency, usually exhibit nonlinear features. Therefore, we can extend the Engle-Granger cointegration test by adjusting the error correction model and applying nonlinear unit root tests to increase the reliability of the study. These extensions have been studied by \textcite{enders2001cointegration}, \textcite{hansen2002testing}, and \textcite{kapetanios2006testing}. 
        Several research papers have been published that specifically explore the application of pairs trading in the cryptocurrency market. These include studies by \textcite{van2018cointegration}, \textcite{pritchard2018digital}, 
        \textcite{kakushadze2019altcoin}, \textcite{leung2019constructing}, and \textcite{tadi2021evaluation}.

        In addition to the commonly used methods discussed above, more advanced concepts such as copulas can be applied to enhance the empirical results of the strategy. Compared to correlation or linear cointegration-based methods, the copula approach provides more valuable information regarding the shape and characteristics of pairs' dependency \parencite{ferreira2008new}. \textcite{xie2013copula} demonstrated that the two commonly used pairs trading methods, namely the distance and cointegration methods, can be generalized as special cases of the copula method under certain conditions, and the dependency structure of assets in the copula approach is more robust and accurate.
        
        Moreover, \textcite{liew2013pairs} conducted a comparative study of a copula-based pairs trading strategy with other conventional approaches such as the distance and cointegration approach. They found that the copula approach for pairs trading has better empirical results than the others, as it provides more trading possibilities with higher confidence in practice and does not entail any rigid assumptions, such as the linearity association between assets' returns, in traditional approaches. Hence, the copula approach can provide closer estimations and predictions of reality. According to \textcite{stander2013trading}, the copula approach can significantly demonstrate the pairs' dependency, as it can capture the asymmetry and heavy-tail characteristics of asset returns to model the marginal distribution functions instead of modeling them by Gaussian distribution. Additionally, their empirical analyses reveal that there are more trading opportunities when the market is highly volatile, and the profitability of the strategy strongly depends on the liquidity of the market.
        
        According to \textcite{krauss2017non}, the copula approach can be divided into two sub-streams: return-based and level-based. In the return-based copula method, the log-returns of two assets are calculated, and their marginal distributions are estimated. Then, an appropriate copula is chosen to represent the dependency relationship between the two assets. To generate trading signals, a mispricing index is defined to indicate the degree of abnormal relationship between the assets. Unlike the distance and cointegration methods which use spread-based mispricing definition, the copula method defines mispricing based on the copula's conditional probability distribution of the corresponding assets' log returns. The conditional distribution of copulas can be obtained by taking the partial derivative of the copula function, as shown below\footnote{For more details, see Section 4.2.}:
            \begin{equation}\label{eq conditional copula}
                \begin{split}
                    h^{1|2}:=h(u_1|u_2)=P(U_1\leq u_1|U_2=u_2)=\frac{\partial C(u_1,u_2)}{\partial u_2}\\
                    h^{2|1}:=h(u_2|u_1)=P(U_2\leq u_2|U_1=u_1)=\frac{\partial C(u_1,u_2)}{\partial u_1}
                \end{split}
    		\end{equation}
        where $C(u_1,u_2)$ is the copula distribution function, $h^{1|2}$ and $h^{2|1}$ are conditional copula distribution functions, and $U_1$ and $U_2$ are the transformed uniform variables of the log returns. The values of $h^{1|2}$ and $h^{2|1}$ are between $0$ and $1$, and as much as their values get away from $0.5$, we can consider it as a deviance from the expected relationship between the two assets. \textcite{ferreira2008new}, \textcite{liew2013pairs}, \textcite{stander2013trading}, and \textcite{keshavarz2023profitability} deployed their strategy in that way.
        
        The usage of the return-based method has a drawback in that entry and exit signals are linked with only the previous period's return. To address this limitation, a new approach (hereafter referred to as the value-based method) was proposed by \textcite{xie2013copula}. They defined a new mispricing index by aggregating the surplus value of conditional probability in equation \ref{eq conditional copula} from 0.5 across multiple periods to determine the extent to which assets are out of balance. 
        \begin{equation}\label{eq CMI}
        	        \begin{split}
        	        \text{CMI}^{1|2}_{t} = \text{CMI}^{1|2}_{t-1} + \left({h}_{t}^{1|2} - 0.5\right) \\
        	    \text{CMI}^{2|1}_{t} = \text{CMI}^{2|1}_{t-1} + \left({h}_{t}^{2|1} - 0.5\right)
        	        \end{split}
        \end{equation}
        The studies conducted by \textcite{xie2016pairs},  \textcite{rad2016profitability}, \textcite{krauss2017non}, and \textcite{da2023pairs}. According to \textcite{xie2016pairs}, the copula method outperforms the distance approach in terms of describing the dependency relationship between assets and identifying more statistical arbitrage opportunities that generate greater profits. They also recommended utilizing the copula approach for high-frequency pairs trading. \textcite{rad2016profitability} compared the performance of the copula approach with that of traditional methods using daily US stocks. They found that while the profitability of the copula method could be weaker, it is more reliable in capturing arbitrage opportunities in the US stock market. They determined that the Student-t copula is more appropriate for modeling the dependence structure of pairs in the US stock market than other copulas.
    
        The performance of a pairs trading strategy based on a weighted combination of copulas was evaluated by \textcite{da2023pairs} against a distance methodology, using a vast dataset of $S\&P500$ stocks over 25 years. The study examined the effect of financial factors on profitability. The mixed copula approach yielded better results than the distance method, with higher alphas for fully invested capital and superior performances overall. The approach was notably effective under committed capital during both crisis and non-crisis periods and fully invested during non-crisis periods.

        However, in practice, the cumulative mispricing indices (CMI) in equation \ref{eq CMI} may not necessarily exhibit mean-reverting behavior, which can negatively impact the profitability of the strategy. Hence, a new methodology is proposed in this study to replace log-returns with stationary spread processes in order to overcome this issue\footnote{For more details, see Section 5.}.
    
    \section{Cryptocurrency Exchange and Data Source}\label{ch 3}
        The cryptocurrency market facilitates decentralized trading of cryptocurrencies across various exchanges. Binance, established in 2017, is the largest cryptocurrency exchange worldwide in terms of daily trading volume in both spot and derivatives markets. Binance provides three types of derivative contracts: Futures contracts, Options, and Binance Leveraged Tokens (BLVT). 

        Futures contracts are classified into two main categories: COIN-Margined contracts and USD\textcircled{s}-Margined contracts. COIN-Margined contracts are inverse Futures quoted in US dollars but denominated in an underlying cryptocurrency (e.g., Bitcoin). They include traditional quarterly Futures and perpetual Futures, also known as perpetual Swaps, which never expire or settle. Because they lack a settlement price, their price can deviate significantly from their spot contract price. Binance uses funding fees to address this issue for long and short sides. On the other hand, USD\textcircled{s}-Margined contracts are similar to COIN-Margined Futures, and they may have perpetual or quarterly expiration. However, they are quoted, denominated, and settled in stablecoins such as Tether (USDT) and Binance USD (BUSD), which are pegged to the US dollar value. For this study, all nominated cryptocurrency coins are USDT-Margined Futures \parencite{Binance}.

        The minimum trade amount, also known as the minimum price increment, is the smallest possible change in the price of a contract at an exchange. This value is specific to each asset and can be adjusted over time. Assets with smaller increments have narrower bid/ask spreads. When limit orders are not entirely disclosed to traders, they tend to order contracts with smaller quote sizes to avoid slippage \parencite{harris1997decimalization}.

        To calculate profit and loss, all coins are valued in Tether (USDT), a stablecoin. Using the Binance API, we collected historical hourly closed prices for twenty cryptocurrency coins from 01/01/2021 to 19/01/2023. We used the linear Engle-Granger (EG) two-step method and to identify cointegrated pairs and calculated Kendall's Tau. Then, we calibrated the method's parameters using the copula approach and generated trading signals for the strategy.

        To calculate profit and loss, all coins are valued in Tether (USDT), a stablecoin. Using the Binance API, we collected historical hourly closed prices for twenty cryptocurrency coins from 01/01/2021 to 19/01/2023. We used Engle-Granger (EG) and Kapetanios-Shin-Shell (KSS) cointegration tests to identify cointegrated pairs and calculated Kendall's Tau. Then, we calibrated the method's parameters using the copula approach and generated trading signals for the strategy.

        A pairs trading cycle comprises formation and trading periods. In this study, the entire process of pairs trading is conducted within a month, with three weeks allocated for the formation step and the remaining week designated for the trading step. We carry out 104 cycles that move dynamically over time, with each cycle sharing three-quarters of its data with its previous or subsequent cycle. 
        \begin{figure}[H]
            \caption {\small \textbf{The scheme of Formation and Trading Periods}}
            \includegraphics[scale=0.7]{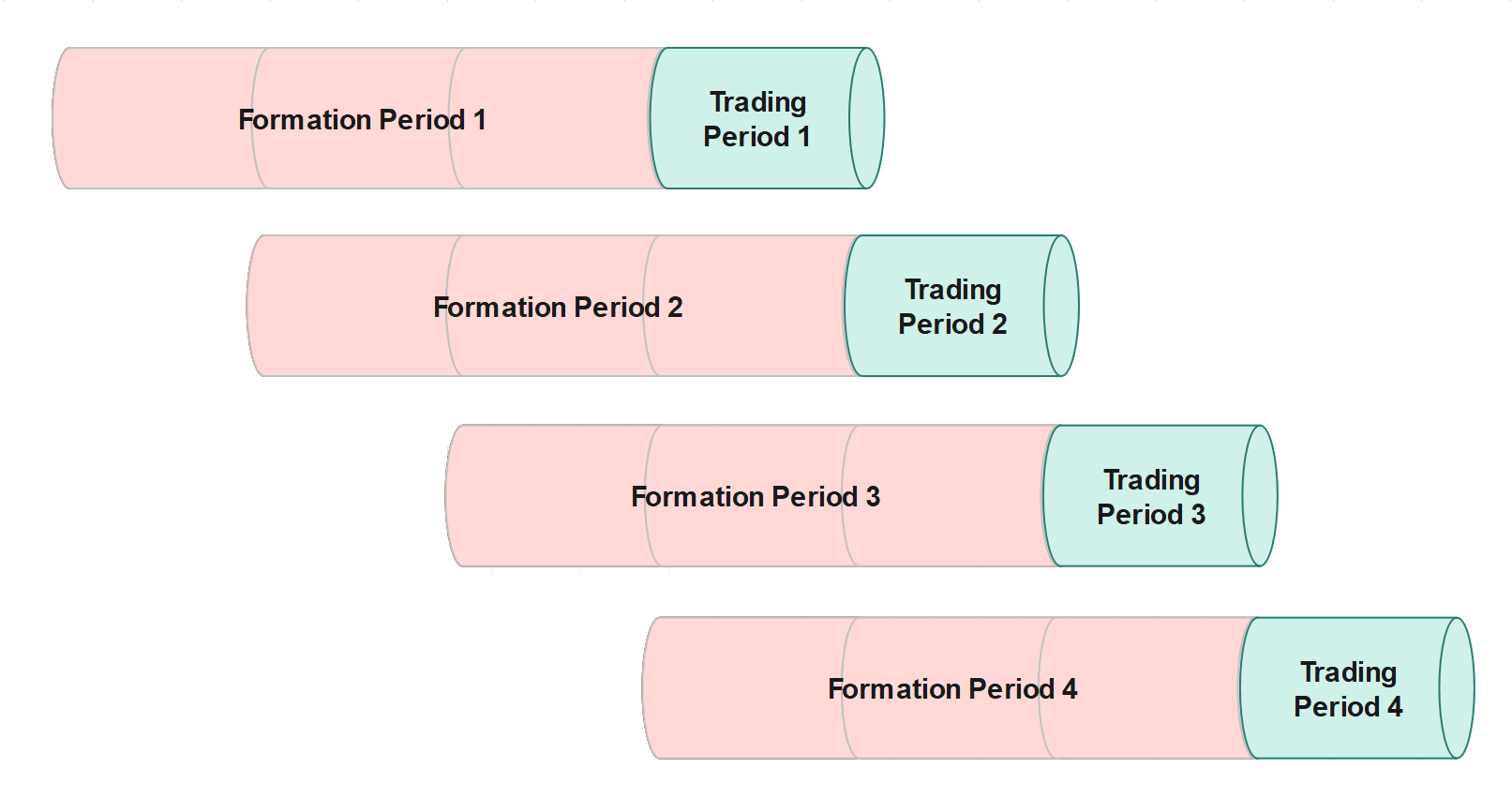}
            \centering
            \label{fig. 1}
        \end{figure}
        Furthermore, we implemented the entire methodology using Python and R, with Python being the preferred option for data handling and R for statistical tests and copula modeling.
        \begin{table}[H]
            \caption{\small \textbf{Python libraries and R packages used in this research}}

            \centering
            \begin{tabular}{@{}llll@{}}
                \toprule
                \multicolumn{2}{c}{\textbf{Python}}                         & \multicolumn{2}{c}{\textbf{R}}                               \\ 
                \midrule
                \multicolumn{1}{c}{\textbf{Name}} & \multicolumn{1}{c}{\textbf{Type}} & \multicolumn{1}{c}{\textbf{Name}} & \multicolumn{1}{c}{\textbf{Type}} \\
                \midrule
                NumPy                             & Numerical analysis                & copula                            & Copula modeling                   \\
                Pandas                            & Technical computing               & VineCopula                        & Copula modeling                   \\
                Matplotlib                        & Plotting                          & MASS                      & Statistical package               \\
                Datetime                          & Date/time package                 & dpylr                             & Data handling                     \\
                Statsmodels                       & Statistical package               & lubridate                         & Date/time package                 \\
                SciPy                             & Technical computing               & fUnitRoots                     & Unit root analysis                         \\ \bottomrule
                \end{tabular}
            \label{tab 2}
        \end{table}
    
    \section{Theoretical Framework}	\label{ch 4}
        \subsection{Unit-Root Test}
            The cointegration property can be employed to identify the most appropriate pair from multiple combinations of coins. Both linear and non-linear cointegration tests can be utilized for evaluation. Initially, the pair spread value is defined without an intercept in the following manner:
            \begin{equation}\label{eq 4.1}
            S_t = P^1_t - \beta P^2_t
            \end{equation}
            Suppose that $P^1_t$ and $P^2_t$ are non-stationary time series. We use unit-root tests to study whether the spread, $S_t$, is also a non-stationary process or not. The augmented Dickey-Fuller (ADF) unit-root test, as described by Dickey and Fuller (1979), uses a test equation applied to the demeaned spread process $S_t$ in the following form:
            \begin{equation}\label{eq 4.3}
                \Delta S_t = \beta S_{t-1} + \sum_{i=1}^{p-1}\gamma_i\Delta S_{t-i} + \epsilon_t,
            \end{equation}
            where $\beta$ is the coefficient of the lagged level of the series, $\gamma_i$ are the coefficients of the lagged differences, $\epsilon_t$ is the error term, and $p$ is the number of lags used in the test. The null hypothesis of the ADF test is that the series has a unit root, i.e., $\beta=0$. If the test statistic exceeds a critical value at a given significance level, the null hypothesis is rejected, indicating that the series is stationary and does not have a unit root. Conversely, if the test statistic is below the critical value, the null hypothesis cannot be rejected, implying that the series has a unit root and is non-stationary.

            Traditional unit-root tests like the Augmented Dickey-Fuller (ADF) test assume that the data-generating process is linear. However, non-linear unit-root tests are designed to account for non-linearities in time series data and provide more accurate assessments of unit roots. There are several non-linear unit-root tests available in the literature, each with its own assumptions, methodologies, and advantages. Examples include the \textcite{terasvirta1994specification} test, the \textcite{zivot2002further} test, the \textcite{kapetanios2003testing} test, and the \textcite{kapetanios2005unit} test. These tests often involve estimating non-linear models like threshold auto-regressive (TAR) models, smooth transition auto-regressive (STAR) models, or other non-linear models, and computing test statistics to compare estimated model parameters with critical values.
            
            The general self-exciting threshold auto-regressive (SETAR) model with $n$ regimes applied to the demeaned spread process $S_t$ is in the following form:
            \begin{equation}\label{eq 4.4}
                S_t = \sum_{i=1}^p \left(\phi_{i1} \mathbbm{1}_{\{S_{t-d} \leq c_1\}} + \sum_{j=1}^{n-1} \phi_{ij} \mathbbm{1}_{\{c_j < S_{t-d} \leq c_{j+1}\}}+\phi_{in} \mathbbm{1}_{\{S_{t-d} > c_n\}}\right)S_{t-i} + \epsilon_t,
            \end{equation}
             where $d$ denotes the transition's delay, $c_j$ represents the $j$-th threshold, and $\epsilon_t$ represents the error term. In special case \textcite{kapetanios2003testing} proposed a test equation where the indicator function is replaced by an exponential smooth transition function in the form:
            \begin{equation}\label{eq 4.7}
            S_t = S_{t-1}+\sum_{i=1}^p\left(\gamma_{1i}\,\left(1-e^{-\theta \left(S_{t-1}-c\right)^2}\right)\,S_{t-i}\right)+ \epsilon_t
            \end{equation}
            When $c$ is set to zero, and $p$ is set to one, using Taylor approximation, equation \ref{eq 4.7} can be illustrated as
            \begin{equation}\label{eq 4.10}
                \Delta S_t = \delta \left(S_{t-1}\right)^3+\epsilon^{'}_t
            \end{equation}
            where $\delta = \gamma_{1}\theta$ and $\epsilon^{'}_t = f(\epsilon_{t})$. The null hypothesis assumes that $\delta$ is equal to zero, while the alternative hypothesis posits that $\delta$ is less than zero. Note that the asymptotic standard normal distribution of the t-statistic for $\delta=0$ against $\delta<0$ is not applicable. However, its asymptotic critical values have been determined through stochastic simulations and are documented by \textcite{kapetanios2003testing}.
        \subsection{Copula Concept}
            Suppose that a continuous random variable $X$ has a continuous distribution, and its probability distribution function is defined as $F_X(x):=\mathbb{P}(X\leq x)$. If $F_X$ is strictly increasing, then $F_X^{-1}$ is defined by $F_X^{-1}(u) = x \Leftrightarrow F_X(x)=u$. However, if $F_X$ is constant on some interval, then the inverse function is not well defined by $F_X^{-1}(u) = x$. To avoid this problem, we can define $F_X^{-1}(u)$  for $0<u<1$ by the generalized inverse function such that
            \begin{equation}\label{eq 4.11}
        	    F_X^{-1}(u) = \inf\{x:F_X(x) \geq u\},
            \end {equation}
            Now, in the same way, we define another continuous random variable Y with distribution function $F_Y$ and generalized inverse function $F_Y^{-1}$ similar to equation \ref{eq 4.11}. Given two continuous random variables $X$ and $Y$, with distribution functions $F_X$ and $F_Y$ respectively, the joint distribution function $F_{X,Y}$ can be written as:
            \begin{equation}
            F_{X,Y}(x,y) = \mathbb{P}(X \leq x, Y \leq y) = \mathbb{P}(F_X(X) \leq F_X(x), F_Y(Y) \leq F_Y(y))
            \end{equation}
            where the last equality follows from the fact that $F_X$ and $F_Y$ are both increasing. Then we define random variables $U$ and $V$ such that $U:= F_X(X)$ and $V:=F_Y(Y)$. According to the probability integral transformation theorem, the probability distribution function of $U$, $F_U$, and the probability distribution function of $V$, $F_V$, are uniformly distributed on $[0,1]$. (See \textcite[p.~54-55]{casella2021statistical} for the proof). 
            
            \textbf{Definition:} A two-dimensional copula $C$ is a function that maps the unit square $[0,1]^2$ into the unit interval $[0,1]$, satisfying the following requirements:
            \begin{enumerate}
                \item $C(0,v)=C(u,0)=0$,\, for $0\leq u,v\leq 1$.
                \item $C(u,1)=u$, and $C(1,v)=v$, \, for $0\leq u,v\leq 1$.
                \item $C(u_1,v_1)-C(u_1,v_2)-C(u_2,v_1)+C(u_2,v_2) \geq 0$, \,  for $1 \geq u_1>u_2\geq 0$, and $1 \geq v_1>v_2\geq 0$.
            \end{enumerate}
            We can define several copula functions, but the three requirements above should be satisfied by $C(u,v)$ to have a well-defined joint distribution function \parencite{cherubini2011dynamic}.
            According to Sklar's theorem, there exists a Copula function $C$ which could connect the uniform random variables $U$ and $V$ to the joint distribution function $F_{X, Y}$ as follows
            \begin{equation}\label{eq 4.13}
                F_{X,Y}(X,Y)=F_{X,Y}(F_X^{-1}(U),F_Y^{-1}(V)):=C(U,V),
            \end{equation}
            Hence, we can rewrite the joint distribution function of X and Y in terms of standard uniform random variables U and V such that
            \begin{equation}\label{eq 4.14}
                F_{X,Y}(x,y) = F_{X,Y}(F_X^{-1}(u),F_Y^{-1}(v)):=C(u,v)=C(F_X(x),F_Y(y)),
            \end{equation}
            where $u=F_X(x)$, and $v=F_Y(y)$. Knowing $F_{X,Y}(x,y)=C(u,v)$, we can determine the copula density function $c(u,v)$ by
            \begin{equation}\label{eq 4.15}
                c(u,v)=\frac{\partial^2C(u,v)}{\partial u\, \partial v}
                =\frac{\partial^2F_{X,Y}(x,y)}{\partial F_X(x)\,\partial F_Y(y)}
                =\frac{\frac{\partial^2F_{X,Y}(x,y)}{\partial x\,\partial y}}{\frac{\partial F_X(x)\,\partial F_Y(y)}{\partial x\,\partial y}}
                =\frac{f_{X,Y}(x,y)}{f_{X}(x)f_{Y}(y)}
            \end{equation} 
            Sklar's theorem enables us to separate the modeling of the marginal distributions $F_X(x)$ and $F_Y(y)$ from the dependence relation represented in C. 
            We can define  the conditional distribution functions using copula function. The conditional distribution of $Y|X=x$ is obtained by the first partial derivatives of the copula function as follows
            \begin{equation}\label{eq 4.16}
            \begin{split}                
                F_{Y|X}(y)&= \mathbb{P}(Y\leq y|X=x)
                =\mathbb{P}(F_Y(Y)\leq F_Y(y)|F_X(X)=F_X(x))\\
                &=\mathbb{P}(V\leq v|U=u)
                =\lim_{\delta \to 0^{+}} \frac{\mathbb{P}(V\leq v, U\in (u-\delta, u+\delta)}{\mathbb{P}(U\in (u-\delta, u+\delta))}\\
                &=\lim_{\delta \to 0^{+}} \frac{C(u+\delta,v) - C(u-\delta,v)}{2\delta}\\
                &=\frac{\partial}{\partial u}C(u,v).
            \end{split}
            \end{equation}
            Copulas are invariant concerning strictly increasing transformations of the marginal distributions. For more details about the characterization of invariant copulas, see \textcite{klement2002invariant}. Also, we can increase the range of dependence captured by copulas by rotating them. The following equations show how to rotate a copula by 90, 180, and 270 degrees:
            \begin{equation}
                \begin{split}
                &C_{90}(u_1,u_2) := C(1-u_2, u_1)\\
                &C_{180}(u_1,u_2) := C(1-u_1, 1-u_2)\\
                &C_{270}(u_1,u_2) := C(u_2, 1-u_1)
                \end{split}
            \end{equation}
            \subsection{Families of copulas}
            There are several types of copulas. This research focuses on three popular families of copulas: elliptical, Archimedean, and extreme value copulas. Each type of copula is characterized by its properties, such as its dependence structure and tail behavior, and is often used in different areas of statistics and finance.
            \subsubsection{Elliptical Copulas}
            Elliptical copulas are widely used in finance and insurance applications due to their ability to model dependence with different levels of tail dependence. An elliptical copula is constructed from a multivariate elliptical distribution. The most commonly used elliptical distributions are the multivariate Gaussian and Student-t. The density function of any elliptical distribution $f_{\boldsymbol{X}}$ is in the form
            \begin{equation} \label{eq 4.17}
                f_{\boldsymbol{X}}(\boldsymbol{x};\boldsymbol{\mu},\boldsymbol{\Sigma})=k_n|\boldsymbol{\Sigma}|^{-\frac{1}{2}}g\left(\left(\boldsymbol{x} - \boldsymbol{\mu})^T\boldsymbol{\Sigma}^{-1}(\boldsymbol{x} - \boldsymbol{\mu}\right)\right),
            \end{equation}
            where $k_n \in \mathbb{R}$ is the normalizing constant and dependents on the dimension $n$, $\boldsymbol{x}$ is an n-dimensional random vector with mean vector $\boldsymbol{\mu}$, and a positive definite matrix which is proportional to the covariance matrix $\boldsymbol{\Sigma}$, and some function $g(.)$ which is independent of the dimension $n$ \parencite{czado2019analyzing}. In the the case of bivariate Gaussian distribution $g(x):=e^{-x/2}$, $k_n:=1/(2\pi)$, and the probability density function of $\boldsymbol{X}=(X_1,X_2)$ is
            \begin{equation}\label{eq 4.18}
               \begin{split}
                   {f_{\boldsymbol{X}}(\boldsymbol{x};\boldsymbol{\mu},\boldsymbol{\Sigma})}
                   =\frac {1}{2\pi|\boldsymbol{\Sigma}|^{1/2}}\mathrm {e} ^{-{\frac{1}{2}}\left[\left({ {\boldsymbol{x}-\boldsymbol{\mu}}}\right)^T\boldsymbol{\Sigma}^{-1}{(\boldsymbol{x}-\boldsymbol{\mu})}\right]}\\
                    {\boldsymbol{\mu} }={\begin{pmatrix}\mu_1\\\mu_2\end{pmatrix}},\quad {\boldsymbol\Sigma }={\begin{pmatrix}\sigma_1^{2}&\rho \sigma_1\sigma_2\\\rho \sigma_1\sigma_2&\sigma_2^{2}\end{pmatrix}},
               \end{split}
            \end{equation}
            where $\rho$ is the correlation between random variables $X_1$ and $X_2$, $\sigma_1>0$ and $\sigma_2>0$. If $\mu_1=\mu_2=0$ and $\sigma_1=\sigma_2=1$, then the density and distribution functions of the standard Bivariate Gaussian distribution are obtained by
            \begin{equation}\label{eq 4.19}
                \begin{split}
                    &\phi_{X_1X_2}(x_1,x_2;\rho) = \frac {1}{2\pi {\sqrt {1-\rho ^{2}}}}{e} ^{-{\frac {x_1^2 -2\rho x_1 x_2 +x_2^2}{2(1-\rho ^{2})}}}\\
                    &\Phi_{X_1X_2}(u_1,u_2;\rho) = \int_{-\infty}^{u_1}\int_{-\infty}^{u_2}\phi_{X_1X_2}(x_1,x_2;\rho)\,dx_1\,dx_2
                \end{split}
            \end{equation}
            Using Sklar's theorem in \ref{eq 4.13} and \ref{eq 4.19}, the bivariate Gaussian copula is defined by
            \begin{equation}\label{eq 4.20}
               C(u_1, u_2;\rho) := \Phi_{X_1X_2}(\Phi_{X_1}^{-1}(u_1), \Phi_{X_2}^{-1}(u_2);\rho) 
            \end{equation}
            In the bivariate t distribution, $g(.)$ and $k_n$ function in the formula \ref{eq 4.17} are defined by $g(x):=(1+x/\nu)^{-(\nu+2)/2}$, $k_n:=\Gamma(\frac{\nu+2}{2})/\left(\Gamma(\frac{\nu}{2})\nu\pi\right)$, and the probability density function of $\boldsymbol{T}=(T_1,T_2)$ is obtained by
            \begin{equation}\label{eq 4.21}
                \begin{split}
                    &f_{\boldsymbol{T}}(\boldsymbol{t};\nu,\boldsymbol{\mu},\boldsymbol{\Sigma})
                    =\frac{\Gamma\left(\frac{\nu + 2}{2}\right)}{\Gamma\left(\frac{\nu}{2}\right)\nu\pi|\boldsymbol{\Sigma}|^{1/2}}\left[1+\frac{1}{\nu}(\boldsymbol{t}-\boldsymbol{\mu})^T\boldsymbol{\Sigma}^{-1}(\boldsymbol{t}-\boldsymbol{\mu})\right]^{-(\nu+2)/2}\\
                \end{split}
            \end{equation}
            where $\nu>0$ is the degree of freedom parameter and $ \boldsymbol{\mu}$ and $\boldsymbol{\sigma}$ are the same as \ref{eq 4.18}. If $\mu_1=\mu_2=0$ and $\sigma_1=\sigma_2=1$, and knowing that ${ {\Gamma \left({\frac {\nu +2}{2}}\right)}/{\Gamma \left({\frac {\nu }{2}}\right)}}={\frac {\nu}{2}}$, the density and distribution function of the standard Bivariate Student-t distribution are obtained by 
            \begin{equation}\label{eq 4.22}
                \begin{split}
                    &f_{T_1T_2}(t_1,t_2;\nu,\rho)=\frac{(1-\rho^2)^{-1/2}}{2\pi}\left[1+\frac{t_1^2-2\rho t_1t_2+t_2^2}{\nu(1-\rho^2)}\right]^{-(\nu+2)/2}\\
                    &F_{T_1T_2}(u_1,u_2;\nu,\rho)=\int_{-\infty}^{u_1}\int_{-\infty}^{u_2}f_{T_1T_2}(t_1,t_2;\nu,\rho)\,dt_1\,dt_2
                \end{split}
            \end{equation}
            Then the bivariate Student-t copula is defined by
            \begin{equation}\label{eq 4.23}
                C(u_1, u_2;\nu,\rho) := F_{T_1T_2}(F_{T_1}^{-1}(u_1), F_{T_2}^{-1}(u_2);\nu,\rho) 
            \end{equation}
            \subsubsection{Archimedean Copulas}
            Archimedean copulas are based on a generator function and can model dependence with tail dependence that decreases logarithmically or exponentially. According to \textcite{nelsen2007introduction}, Archimedean copulas are defined as follows:
            \begin{equation}\label{eq 4.24}
                C(u_1,\dots , u_n) = \phi^{[-1]}(\phi(u_1)+\dots+\phi(u_n))
            \end{equation}
            where $\phi: [0,1] \rightarrow [0,\infty]$ is called a generator which is continuous, strictly decreasing, and convex function such that $\phi(1) = 0$. In addition, $\phi^{[-1]}: [0,\infty] \rightarrow [0,1]$ is called the pseudo-inverse of $\phi$ function and is defined by
            \begin{equation}\label{eq 4.25}
                \phi^{[-1]}(t) = 
                \begin{cases} 
                     \phi^{-1}(t) & ,\,0\leq t\leq  \phi(0) \\
                    0 & ,\,\phi(0) \leq t\leq  \infty
                \end{cases}
            \end{equation}
            If $\phi(0)$ equals infinity, then the pseudo-inverse function $\phi^{[-1]}$ is equivalent to the inverse function $\phi^{-1}$. Archimedean copulas allow for a range of generator functions to be selected, which ultimately determines the type of tail dependence exhibited by the copula. For this research, both one-parameter copulas (such as the Gumbel, Clayton, Frank, and Joe copulas) and two-parameter copulas (such as BB1, BB6, BB7, and BB8) were utilized. Table \ref{tab 3} in the appendix provides a list of the bivariate Archimedean copulas utilized in this study.
            
        \subsubsection{Extreme-Value Copulas}
            Extreme value copulas are used to model dependence with strong tail dependence, making them suitable for modeling extreme events. Suppose that $X_i = \left(X_{i1}, X_{i2}\right)^T$, $i \in \{1,2,\cdots,n\}$, be independent and identically distributed random vectors with joint distribution function $F$, and marginal distributions $F_1$ and $F_2$. According to \textcite{gudendorf2010extreme}, we can define the bivariate vector of component-wise maxima $M_n = (M_{n1}, M_{n2})^T$ such that 
            \begin{equation}\label{eq 4.26}
                M_{nj} := \max_{i \in \{1,2,\cdots,n\}}\left(X_{ij}\right),\quad j = 1,2.
            \end{equation}
            Then, the bivariate copula $C_n$ of $M_n$ is obtained by
            \begin{equation}\label{eq 4.27}
                C_n(u_1,u_2) = C_F\left(u_1^{1/n},u_2^{1/n}\right)^n,\quad (u_1,u_2) \in [0,1]^2.
            \end{equation}
            In equation \ref{eq 4.27}, if $C_F$  exists such that
            \begin{equation} \label{eq 4.28}
                \lim_{n\rightarrow \infty}C_F (u_1^{1/n}, u_2^{1/n})^{n} = C(u_1,u_2), \quad (u_1,u_2) \in [0,1],
            \end{equation}
            then the bivariate copula $C$ in (\ref{eq 4.28}) is called an extreme-value copula. Bivariate extreme-value copulas can be demonstrated in terms of a function $A(t)$ in this form:
            \begin{equation}\label{eq 4.29}
                C(u_1,u_2) = (u_1u_2)^{A\left(\ln(u_2)/\ln(u_1u_2)\right)},\quad (u_1,u_2) \in (0,1]^2\setminus \{(1,1)\},
            \end{equation}
            where the function $A: [0,1] \rightarrow [1/2,1]$, which is called the Pickands dependence function, is convex and satisfies $\max(1-t, t) \leq A(t) \leq 1$ for all $t \in [0, 1]$ \parencite{gudendorf2010extreme}. Some Archimedean copulas, such as the Gumbel copula, can be expressed by the extreme-value family. These special copulas create a hybrid category, including both the Archimedean and the extreme-value copulas, and are called Archimax copulas.\par
            Table \ref{tab 4} shows extreme-value copulas used in this research. Note that Tawn copula has three parameters and its Pickands's dependence function is in the form
            \begin{equation}\label{eq 4.30}
                A(t) = (1-\beta)+(\beta-\alpha)t+\left[(\alpha(1-t))^\theta + (\beta t){\theta}\right]^{1/\theta},
            \end{equation} where $\theta \geq 1$ and $\alpha,\beta \in [0,1]1$. The simplified Tawn copula cases with $\beta = 1$ and $\alpha = 1$ are respectively called Tawn Type 1 and Tawn Type 2 copula and have two parameters.
            \begin{table}[H]
                \caption{\small \textbf{Pickands dependence function of some extreme-value copulas}}
              \centering
                \begin{tabular}{lll}
                \toprule
                \textbf{Name} & \multicolumn{1}{l}{\textbf{Pickands function $\textbf{A(t)}$}} & \multicolumn{1}{l}{\textbf{Parameters}} \\
                \midrule
                Gumbel &   $\bigg[t^{\theta} + (1-t)^{\theta}\bigg]^{1/\theta}$    &  $\theta \geq 1$ \\
                Tawn Type 1 &   $(1-\alpha)t + \bigg[\left(\alpha(1-t)\right)^{\theta} + t^{\theta}\bigg]^{1/\theta}$  &  $\theta \geq 1, \,\,0 \leq \alpha \leq 1$  \\
                Tawn Type 2 &  $(1-\beta)(1-t)\bigg[(1-t)^\theta + (\beta t){\theta}\bigg]^{1/\theta}$     & $\theta \geq 1, \,\,0 \leq \beta \leq 1$ \\\bottomrule
                \end{tabular}%
              
              \label{tab 4}%
            \end{table}%
        \subsection{Copula Estimation}
            When the marginal probability density of $X_1$ and $X_2$ and their corresponding copula density $c(.)$ are given in their parametric with unknown parameters, we can estimate the vector of parameter vector $\theta= (\beta,\alpha)^T$ with the maximum likelihood estimation method, where $\beta=(\beta_1,\beta_2)^T$ represent the marginal parameters and $\alpha$ represents the copula parameters. The log-likelihood function for $(X_1,X_2)$, where $X_i=(x_{i1},\dots,x_{in})$, can be expressed as
            \begin{equation}\label{eq 4.31}
                l(\theta)=\sum_{j=1}^n\bigl[\log c\bigl(F_1(x_{1j};\beta_1),F_2(x_{2j};\beta_2);\alpha\bigr)+\log f_1(x_{1j};\beta_1)+\log f_2(x_{2j};\beta_2)\bigr],
            \end{equation}
            and the maximum likelihood estimator of $\theta$ is  
            \begin{equation}
                \hat{\theta}_{ML}=\underset{\theta}{\text{argmax}}\,l(\theta).
            \end{equation} However, since this approach is computationally expensive, an alternative method called the Inference for the Margins (IFM) two-step method can be used instead. This method is computationally easier to obtain compared to the full maximum likelihood estimation approach. First, we estimate the margins' parameters by performing the estimation of the univariate marginal distributions using the log-likelihood function where the maximum likelihood estimator of $\beta_i$ is
            \begin{equation}\label{eq 4.32}
                \hat{\beta}_{i}=\underset{\beta_i}{\text{argmax}}\,\sum_{j=1}^n\left[\log f_i(x_{ij};\beta_i)\right].
            \end{equation}
            Then given estimated marginal parameters, we transform data to the copula scale, develop the copula model, and estimate the copula parameters $\alpha$ as follows
            \begin{equation}\label{eq 4.33}
                \hat{\alpha}_{ML}=\underset{\alpha}{\text{argmax}}\,\sum_{j=1}^n\left[\log c\left(F_1(x_{1j};\hat{\beta_1}),F_2(x_{2j};\hat{\beta_2});\alpha\right)\right].
            \end{equation}
            We can also employ a semiparametric approach known as canonical maximum likelihood to estimate copula parameters without specifying the marginals. The empirical cumulative distribution function of $X_i=(x_{i1},\dots,x_{in})$ is
             \begin{equation}\label{eq 4.34}
                {\widehat{F^{i}_n}}(t)=\frac{\#\{X_i\le x\}}{n+1}={\frac {1}{n+1}}\sum _{j=1}^{n}\mathbbm {1} _{\{x_{ij}\leq t\}}.
            \end{equation}
            Then, the copula parameters are estimated using maximum likelihood estimation as follows:
            \begin{equation}\label{eq 4.35}
                \hat{\alpha}_{ML}=\underset{\alpha}{\text{argmax}}\,\sum_{j=1}^n\left[\log c\left(\widehat{F^1_n}(x_{1j}),\widehat{F^2_n}(x_{2j});\alpha\right)\right].
            \end{equation}
    \section{Implementation Methodology} \label{ch 5}
        The study suggests a new approach to address the issue of cumulative mispricing indices (CMI) not exhibiting mean-reverting behavior in practice, which can negatively affect the profitability of trading strategies. The proposed methodology involves using stationary spread processes instead of log-returns. In the pairs trading strategy employed in this study, the spread process is defined as a linear combination of two assets, with a reference asset (BTCUSDT) selected and other cryptocurrency coins identified as cointegrated with it using a specific equation.
        \begin{equation}\label{eq 5.1}
            S^{i}_{t} = \text{BTCUSDT}_{t} - \hat{\beta}^{i} P^{i}_{t}\quad i=1,2,\dots,19
        \end{equation}
        where $\hat{\beta}$ is the estimated linear regression coefficient between BTCUSDT and the second coin, chosen from the other 19 coins. This allows us to identify 19 pairs and select the optimal pairs during the formation period for trading in the trading period. To determine the optimal pairs, we use the linear Engle-Granger (EG) two-step method and the non-linear Kapetanios-Shin-Snell (KSS) cointegration tests to identify cointegrated coins. However, since there may be multiple cointegrated pairs, we need to add another criterion to rank them. To do this, we calculate Kendall's Tau ($\tau$), which is a measure of correlation for ranked data and defined
        \begin{equation}
            \tau(S^i,S^j)=\tau_{ij} = \frac{{\text{{Number of concordant pairs}} - \text{{Number of discordant pairs}}}}{{\text{{Total number of pairs}}}}
        \end{equation}
        where 
        \begin{equation*}
            \text{Number of concordant pairs}=\sum_{n=1}^{N-1} \sum_{m=n+1}^{N} \text{{sgn}}(S^i_{[n]} - S^i_{[m]}) \text{{sgn}}(S^j_{[n]} - S^{j}_{[m]}),
        \end{equation*}  
        \begin{equation*}
            \text{Number of discordant pairs}=\sum_{n=1}^{N-1} \sum_{m=n+1}^{N} \text{{sgn}}(S^i_{[n]} - S^i_{[m]}) \text{{sgn}}(S^j_{[m]} - S^{j}_{[n]}),
        \end{equation*}
        and $\text{total number of pairs}={N(N-1)}/{2}$. Here, $N$ is the number of data points, $S^i_{[n]}$ and $S^{j}_{[n]}$ are the rankings of the $n$-th data point in two different variables, and $\text{{sgn}}(x)$ is the sign function, which is $1$ if $x > 0$, $-1$ if $x < 0$, and $0$ if $x = 0$.
        
        After calculating $\tau$ of BTCUSDT with each of the 19 altcoins, we select the two altcoins that have the highest $\tau$ with BTCUSDT and create their corresponding pairs. During the course of one week, the chosen pairs are traded and can be substituted with different pairs at the start of each trading period. Instead of trading a pair of coins, we employ a pair of spreads where each of them contains BTCUSDT. In this way, having a long position in one spread and a short position in the other one implies that BTCUSDT will not be traded at all and it only plays an intermediary role between two other coins.\par
        Now, we estimate the probability distribution function of spread processes (marginals). We fit various distributions such as Gaussian, Student-t, and Cauchy to the data to identify the best-fitting distribution. By employing statistical methods or maximum likelihood estimation, we estimate the specific parameters associated with each distribution. To evaluate the goodness of fit, we calculate the AIC values for the candidate distributions, ultimately selecting the distribution with the lowest AIC value as the best-fitting option. Suppose that ${F}^{i}$ is the fitted cumulative distribution function of the spread process $S^{i}$.
        The Probability Integral Transform is employed to convert the spreads to random variables $\mathbf{U_1} := {F}^{1}(\mathbf{S^{1}})$ and $\mathbf{U_2} := {{F}}^{2}(\mathbf{S^{2}})$ with a standard uniform distribution. The next step is to determine a fitting copula model for $\mathbf{U_1}$ and $\mathbf{U_2}$. We select some potential copulas and estimate the corresponding parameters by the maximum likelihood method. Finally, using the Akaike information criterion (AIC), we distinguish the most fitted copula model.\par
        During the trading period, using the hourly realizations of random variables $U_1$ and $U_2$, which are the transformed values of spread processes, we calculate the copula conditional probabilities $h^{1|2}$ and $h^{2|1}$ defined in equation \ref{eq conditional copula}, for the selected pairs of the week. when $h^{1|2}$ is higher (lower) than 0.5, the first coin can be considered to be overvalued (undervalued) relative to the second one. Similarly, when $h^{2|1}$ is higher (lower) than 0.5, the second coin can be considered to be overvalued (undervalued) relative to the first one. Therefore, we can interpret mispricing as conditional probabilities in equation \ref{eq conditional copula} minus 0.5.
        We denote the trading thresholds to be $\alpha_1$ and $\alpha_2$. We study the optimal triggers via back-testing. The opening and closing signals are generated by the following rules:
        \begin{table}[H]
        \caption{\small \textbf{Trading rules in terms $S^1$ and $S^2$}}
          \centering
            \begin{tabular}{cc}
            \toprule
            \textbf{Trading Rule} & \textbf{Signals}\\
            \midrule
            If $h^{1|2} < \alpha_1$ and $h^{2|1} >  1-\alpha_1$ & open long $S^1$ and short $S^2$ \\
            If $h^{1|2} >  1-\alpha_1$ and $h^{2|1} < \alpha_1$ & open short $S^1$ and,long $S^2$ \\
            If $|h^{1|2}-0.5|< \alpha_2$ and  $|h^{2|1}-0.5|< \alpha_2$ & close both positions\\\bottomrule
            \end{tabular}%
          \label{tab 5}%
        \end{table}%
        
        \begin{table}[H]
            \caption{\small \textbf{Trading rules in terms $P^1$ and $P^2$}}
          \centering
            \begin{tabular}{cc}
            \toprule
            \textbf{Trading Rule} & \textbf{Signals} \\
            \midrule
            If $h^{1|2} < \alpha_1$ and $h^{2|1} >  1-\alpha_1$ & open long $\beta^2\times P^2$ and short $\beta^1 \times P^1$\\
            If $h^{1|2} >  1-\alpha_1$ and $h^{2|1} < \alpha_1$ & open short $\beta^2\times P^2$ and long $\beta^1\times P^1$\\
            If $|h^{1|2}-0.5|< \alpha_2$ and  $|h^{2|1}-0.5|< \alpha_2$ & close both positions \\ \bottomrule
            \end{tabular}%
          \label{tab 6}%
        \end{table}%
        \begin{figure}[H]
            \caption {\small \textbf{Confidence bands of Gumbel copula ($\theta=2$) at $\alpha_1=5\%$ and $\alpha_2 = 10\%$}}
            \includegraphics[scale=0.55]{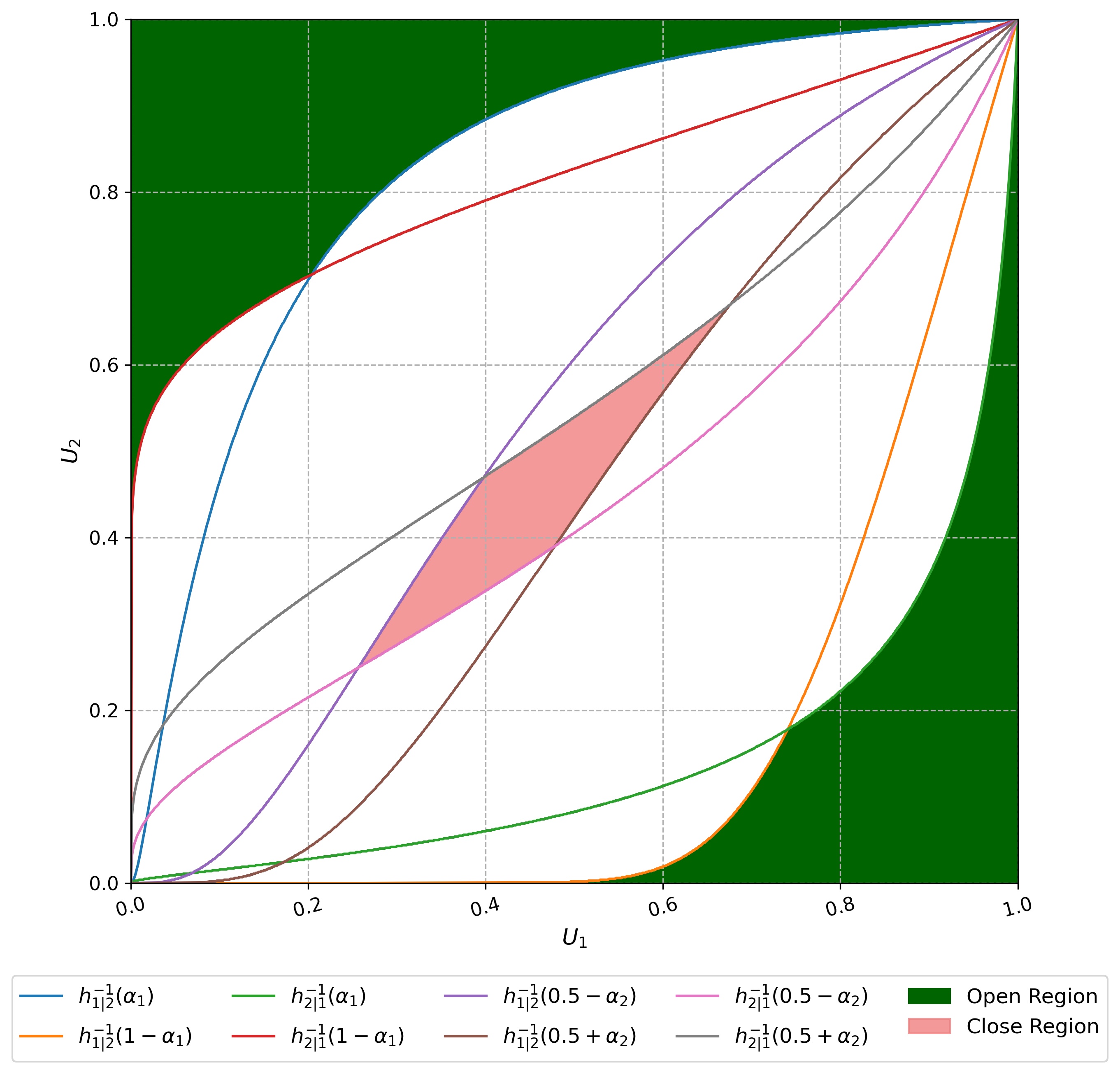}
            \centering
            \label{fig. 2}
        \end{figure}
        Figure \ref{fig. 2} provides an illustrative depiction of the confidence bands in trading rules for the Gumbel copula with a parameter value of $\theta=2$, under the conditions of $\alpha_1=5\%$ and $\alpha_2=10\%$. If the data point $(u_{1t},u_{2t})$ falls within the top green (down green) area, it suggests that $S_1$ is undervalued (overvalued) and $S_2$ is overvalued (undervalued), which may indicate an opportunity to open a position. Conversely, if $(u_{1t},u_{2t})$ falls within the red area, it may signal the need to close the positions.
    \section{Empirical Results}\label{ch 6}
        Table \ref{tab 10} presents the occurrence rate of selected copulas and their rotations over 104 trading weeks. The results indicate that copulas of extreme value, such as Tawn type 1 and 2, and certain two-parameter Archimedean copulas, particularly BB7, play a significant role in the process of selecting the appropriate model.
        \begin{table}[H]
        \caption{\small \textbf{Occurrence rate of copulas in the study}}
          \centering
            \begin{tabular}{lrr}
            \textbf{Copulas and their rotations} & \multicolumn{1}{l}{\textbf{Strategy with EG Test}} & \multicolumn{1}{l}{\textbf{Strategy with KSS Test}} \\
            \toprule
             Gaussian & 4.8\% & 5.8\% \\
             Student-t & 6.7\% & 4.8\% \\
             Clayton & 3.8\% & 6.7\% \\
             Frank & 2.9\% & 3.8\% \\
             Gumbel & 1.0\% & 4.8\% \\
             Joe  & 10.6\% & 7.7\% \\
             BB1  & 5.8\% & 2.9\% \\
             BB6  & 1.9\% & 1.9\% \\
             BB7  & 15.4\% & 15.4\% \\
             BB8  & 8.7\% & 7.7\% \\
             Tawn  type 1 & 23.1\% & 24.0\% \\
             Tawn  type 2 & 15.4\% & 14.4\% \\
            \bottomrule
            \end{tabular}%
          \label{tab 10}%
        \end{table}%
        Tables \ref{tab 7} and \ref{tab 8} in the appendix display the coin pairs chosen for each trading week, which can vary from week to week. It is worth noting that there are instances where the selected coin pairs remain unchanged across consecutive weeks. The ADF unit-root test is conducted with a significance level of $10\%$. For the KSS unit-root test, the asymptotic critical value at a $10\%$ significance level is -1.92. If we fail to identify a minimum of two stationary spread processes within a specified formation period, we abstain from trading during its corresponding trading period. 
        
        The results of our trading strategy's profit and loss calculations with varying entry thresholds ($\alpha_1$) are illustrated in Figure \ref{fig. 2}. Note that the exit threshold ($\alpha_2$) is fixed at $10\%$ in all cases. Realized profit and loss are determined by considering the commission fees and the difference between the opening and closing prices of a position. It should be noted that we initially invested 200,000 USDT, with the coins' weights set to ensure that each side has a maximum initial capital of around 200,000 USDT. It is also assumed that all trades are executed using market orders, which typically incur higher fees (known as taker fees) compared to the lower fees charged for limit orders (known as maker fees). For instance, on Binance, the maker fee of USDT-Margined Futures is set at $0.02\%$, while the taker fee is $0.04\%$. 
        
        \begin{figure}[H]
                \caption {\small \textbf{Copula-based pairs trading strategy P\&L}}
                \includegraphics[scale=0.62]{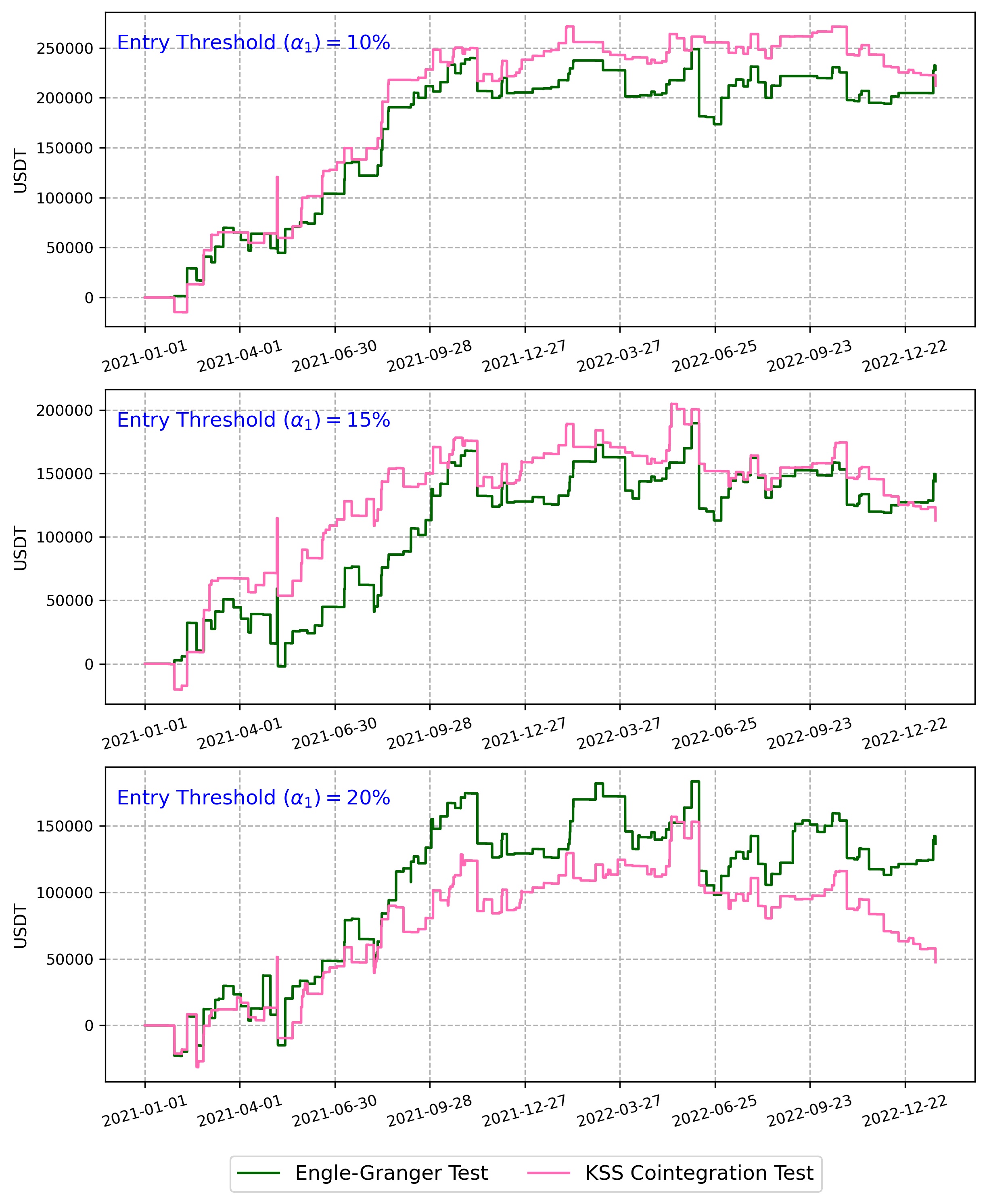}
                \centering
                \label{fig. 3}
        \end{figure}
        
        We assess the risk-adjusted performance of a strategy using the Sharpe ratio. Table \ref{tab 9}\footnote{The Gross Profit and Loss (P\&L) mentioned in table \ref{tab 9} is the total profit and loss generated by transactions before deducting any transaction fees associated with them.\\} illustrates the yearly returns, volatility, and Sharpe ratios for different strategies, along with the maximum drawdown and the returns obtained over the maximum drawdown (RoMaD), which serves as an alternative to the Sharpe Ratio\footnote{Transaction fees are taken into account in all calculations}. In pairs trading with the EG cointegration test, the annualized return decreases with higher entry thresholds, ranging from $76.2\%$ at $\alpha_1 = 0.10$ to $52.1\%$ at $\alpha_1 = 0.20$. In contrast, the annualized standard deviation increases with higher levels of $\alpha_1$, ranging from $38.2\%$ at $\alpha_1 = 0.10$ to $52.7\%$ at $\alpha_1 = 0.20$. In the optimal case, the annualized Sharpe Ratio is approximately 1 ($\alpha_1 = 0.10$). The maximum drawdown increases with higher $\alpha_1$, ranging from $-35.6\%$ at $\alpha_1 = 0.10$ to $-43.9\%$ at $\alpha_1 = 0.20$. The return over maximum drawdown (RoMaD) decreases with higher $\alpha_1$, ranging from $2.14$ at $\alpha_1 = 0.10$ to $1.19$ at $\alpha_1 = 0.20$. 
        
        In a similar manner, the annualized return in pairs trading utilizing the KSS cointegration test diminishes as the entry threshold levels increase, declining from $35.3\%$ when $\alpha_1 = 0.10$ to $10.4\%$ when $\alpha_1 = 0.20$. As $\alpha_1$ increases, the annualized standard deviation increases, starting from $37.3\%$ at $\alpha_1 = 0.10$ and reaching $63.5\%$ at $\alpha_1 = 0.20$. Therefore, the annualized Sharpe Ratio exhibits a declining pattern as the threshold levels increase. It ranges from $0.93$ at $\alpha_1 = 0.10$ to $0.16$ at $\alpha_1 = 0.20$. The maximum drawdown increases with higher threshold levels, ranging from $-34.0\%$ at $\alpha_1 = 0.10$ to $-43.0\%$ at $\alpha_1 = 0.20$. The return over maximum drawdown shows a similar pattern as the annualized Sharpe Ratio, decreasing with higher thresholds.

        When the entry threshold $\alpha_1$ increases, it results in an increase in the number of trading signals. This, in turn, leads to higher volatility in both the EG and KSS strategies. However, despite the increased volatility, there is no corresponding increase in return. This indicates that the optimal threshold is $0.10$, and at this level, the risk-adjusted profitability is similar between the strategy with the EG test and the KSS test.

        To compare our strategy's results with the buy-and-hold strategy, we use a passive investment approach that involves holding a relatively steady portfolio for an extended period, despite short-term market fluctuations. The Bitcoin buy-and-hold strategy shows a negative annualized return of $-17.0\%$ with a high annualized standard deviation of $72.6\%$. The annualized Sharpe Ratio is negative ($-0.23$), indicating poor risk-adjusted performance. The maximum drawdown is severe ($-77.1\%$), indicating a high risk. On the other hand, the portfolio buy-and-hold strategy\footnote{The buy-and-hold strategy in the portfolio involves investing in all twenty cryptocurrency coins with equal weights at the start of the study, retaining them throughout the trading periods, and ultimately selling them at the end of the study period.} shows a positive annualized return of $14.4\%$ with a high annualized standard deviation of $98.6\%$. The annualized Sharpe Ratio is positive ($0.15$), indicating better risk-adjusted performance compared to the Bitcoin buy-and-hold strategy. The maximum drawdown is also severe ($-82.2\%$), indicating a high risk. Both tests of the pairs trading strategy outperform the Bitcoin buy-and-hold strategy and the portfolio buy-and-hold strategy. Our pairs trading strategy outperforms the portfolio buy-and-hold by over six times when $\alpha_1=0.10$.
        \begin{table}[H]
        \caption{\small \textbf{Results of pairs trading strategies Utilizing hourly closed prices of twenty cryptocurrencies from 22/01/2021 to 19/01/2023}}
          \centering
          \resizebox{1\textwidth}{!}{%
            \begin{tabular}{lrrrr}
            \toprule
            \multicolumn{5}{l}{\textbf{Pairs Trading with EG Test}} \\
             &  & $\mathbf{\alpha_1= 0.10}$ & $\mathbf{\alpha_1= 0.15 }$ & $\mathbf{\alpha_1= 0.20}$ \\
                 \midrule
            Total Return &  & \multicolumn{1}{r}{76.2\%} & \multicolumn{1}{r}{54.2\%} & \multicolumn{1}{r}{52.1\%} \\
            Annualized Return &  & \multicolumn{1}{r}{37.1\%} & \multicolumn{1}{r}{26.4\%} & \multicolumn{1}{r}{25.4\%} \\
            Annualized Standard Deviation &  & \multicolumn{1}{r}{38.2\%} & \multicolumn{1}{r}{47.4\%} & \multicolumn{1}{r}{52.7\%} \\
            Annualized Sharpe Ratio &  & \multicolumn{1}{r}{0.97} & \multicolumn{1}{r}{0.56} & \multicolumn{1}{r}{0.48} \\
            Maximum Drawdown &  & \multicolumn{1}{r}{-35.6\%} & \multicolumn{1}{r}{-41.8\%} & \multicolumn{1}{r}{-43.9\%} \\
            Return over Maximum Drawdown &  & \multicolumn{1}{r}{2.14} & \multicolumn{1}{r}{1.30} & \multicolumn{1}{r}{1.19} \\
            Number of Transactions &  & \multicolumn{1}{r}{176} & \multicolumn{1}{r}{200} & \multicolumn{1}{r}{222} \\
            Transaction Costs over Gross P\&L & & \multicolumn{1}{r}{11.7\%} & \multicolumn{1}{r}{19.2\%} & \multicolumn{1}{r}{21.9\%} \\
            \\
            \midrule
            \multicolumn{5}{l}{\textbf{Pairs Trading with KSS Test}} \\
            &  & $\mathbf{\alpha_1= 0.10}$ & $\mathbf{\alpha_1= 0.15 }$ & $\mathbf{\alpha_1= 0.20}$ \\
                 \midrule
            Total Return &  & \multicolumn{1}{r}{72.3\%} & \multicolumn{1}{r}{44.8\%} & \multicolumn{1}{r}{21.3\%} \\
            Annualized Return &  & \multicolumn{1}{r}{35.3\%} & \multicolumn{1}{r}{21.8\%} & \multicolumn{1}{r}{10.4\%} \\
            Annualized Standard Deviation &  & \multicolumn{1}{r}{37.7\%} & \multicolumn{1}{r}{43.6\%} & \multicolumn{1}{r}{63.5\%} \\
            Annualized Sharpe Ratio &  & \multicolumn{1}{r}{0.93} & \multicolumn{1}{r}{0.50} & \multicolumn{1}{r}{0.16} \\
            Maximum Drawdown &  & \multicolumn{1}{r}{-34.0\%} & \multicolumn{1}{r}{-34.6\%} & \multicolumn{1}{r}{-43.0\%} \\
            Return over Maximum Drawdown &  & \multicolumn{1}{r}{2.13} & \multicolumn{1}{r}{1.29} & \multicolumn{1}{r}{0.50} \\
            Number of Transactions &  & \multicolumn{1}{r}{184} & \multicolumn{1}{r}{222} & \multicolumn{1}{r}{254} \\
            Transaction Costs over Gross P\&L &  & \multicolumn{1}{r}{12.9\%} & \multicolumn{1}{r}{25.2\%} & \multicolumn{1}{r}{47.7\%} \\
            \\
            \midrule
            \multicolumn{5}{l}{\textbf{Buy \& Hold Strategy}} \\
            & \multicolumn{2}{r}{Bitcoin Buy \& Hold} & \multicolumn{2}{r}{Portfolio Buy \& Hold} \\
            \midrule
            Total Return & \multicolumn{2}{r}{-33.9\%} & \multicolumn{2}{r}{28.8\%} \\
            Annualized Return & \multicolumn{2}{r}{-17.0\%} & \multicolumn{2}{r}{14.4\%} \\
            Annualized Standard Deviation & \multicolumn{2}{r}{72.6\%} & \multicolumn{2}{r}{98.6\%} \\
            Annualized Sharpe Ratio & \multicolumn{2}{r}{-0.23} & \multicolumn{2}{r}{0.15} \\
            Maximum Drawdown & \multicolumn{2}{r}{-77.1\%} & \multicolumn{2}{r}{-82.2\%} \\
            Return over Maximum Drawdown & \multicolumn{2}{r}{-0.44} & \multicolumn{2}{r}{0.35} \\
            \\
            \bottomrule
            \end{tabular}}%
          \label{tab 9}%
        \end{table}%

    \section{Conclusion}
        In this paper, We develop a novel pairs trading framework for twenty Binance USDT-Margined Futures cryptocurrency coins, which combines copula-based and cointegrated-based approaches. The methodology involves setting a reference asset (in this case, BTCUSDT) and identifying other cryptocurrency coins that are cointegrated with it. To investigate the presence of cointegration, we utilize both the EG two-step method and the KSS cointegration test. After ranking the cointegrated coins based on Kendall's Tau correlation coefficients, we select the two assets with the highest correlation. These selected coins are then traded during the one-week trading period and are updated weekly. Trading rules are generated based on the copula conditional probabilities of the spread processes corresponding to the selected assets. We establish various trading triggers and backtest the strategy.
    
        The research findings show that the choice of an appropriate entry threshold in pairs trading has an important impact on the outcomes. By using lower thresholds, both volatility decreases and return increases concurrently. Our pairs trading strategy has shown a high risk-adjusted return and return over maximum drawdown when compared to the buy-and-hold approach. This suggests that our method can be effectively utilized for profitability. The study also emphasizes the effectiveness of pairs trading in the cryptocurrency market and underscores the importance of meticulous coin pair and copula model selection.
    \printbibliography
    \section*{Appendix}
        \begin{table}[H]
            \caption{\small \textbf{Binance USDT-Margined Futures contracts used in the research}}
          \centering
            \begin{tabular}{llrr}
            \toprule
            \multicolumn{1}{l}{\textbf{Symbol}} & \multicolumn{1}{l}{\textbf{Underlying Crypto}} & \multicolumn{1}{l}{\textbf{Min. Trade Amount}} & \multicolumn{1}{l}{\textbf{Max. Leverage}} \\
            \midrule
            BTCUSDT & Bitcoin & 0.001 BTC & 125x \\
            ETHUSDT & Ethereum & 0.001 ETH & 100x \\
            BCHUSDT & Bitcoin Cash & 0.001 BCH & 75x \\
            XRPUSDT & Ripple & 0.1 XRP & 75x \\
            EOSUSDT & EOS.IO & 0.1 EOS & 75x \\
            LTCUSDT & Litecoin & 0.001 LTC & 75x \\
            TRXUSDT & TRON  & 1 TRX & 50x \\
            ETCUSDT & 	Ethereum Classic & 0.01 ETC & 75x \\
            LINKUSDT & Chainlink & 0.01 LINK & 75x \\
            XLMUSDT & Stellar & 1 XLM & 50x \\
            ADAUSDT & Cardano & 1 ADA & 75x \\
            XMRUSDT & Monero & 0.001 XMR & 50x \\
            DASHUSDT & Dash  & 0.001 DASH & 50x \\
            ZECUSDT & Zcash & 0.001 ZEC & 50x \\
            XTZUSDT & Tezos & 0.1 XTZ & 50x \\
            ATOMUSDT & Cosmos & 0.01 ATOM & 25x \\
            BNBUSDT & Binance Coin & 0.01 BNB & 75x \\
            ONTUSDT & Ontology & 0.1 ONT & 50x \\
            IOTAUSDT & IOTA  & 0.1 IOTA & 25x \\
            BATUSDT & Basic Attention Token & 0.1 BAT & 50x \\\bottomrule 
            \end{tabular}%
          \label{tab 1}%
        \end{table}%
        \begin{table}[H]
            \caption{\small \textbf{Selected pairs using unit-root tests and Kendall's $\tau$ coefficient (part I)}}
          \centering
          \resizebox{0.8\textwidth}{!}{%
            \begin{tabular}{r|lrr|lrr}
            \toprule
                  & \multicolumn{3}{c|}{\textbf{ADF Unit-Root Test Result }} & \multicolumn{3}{c}{\textbf{KSS Unit-Root Test Result}} \\
            \multicolumn{1}{c|}{\textbf{Week}} & \multicolumn{1}{c}{\textbf{Pair}} & \multicolumn{1}{c}{\textbf{P-Value ($S_1$) }} & \multicolumn{1}{c|}{\textbf{P-Value ($S_2$) }} & \multicolumn{1}{c}{\textbf{Pair}} & \multicolumn{1}{c}{\textbf{t-stat ($S_1$) }} & \multicolumn{1}{c}{\textbf{t-stat ($S_2$) }} \\
            \midrule
            1     & ETH-LTC & 0.075 & 0.070 & BCH-ETC & -2.72 & -1.96 \\
            2     & LTC-BCH & 0.023 & 0.010 & LTC-BCH & -3.07 & -2.09 \\
            3     & ETH-LTC & 0.075 & 0.037 & ETH-LTC & -2.10 & -2.90 \\
            4     & ETH-LTC & 0.095 & 0.022 & ETH-ETC & -2.35 & -2.64 \\
            5     & LTC-EOS & 0.096 & 0.082 & LTC-BCH & -2.48 & -2.66 \\
            6     & LINK-TRX & 0.090 & 0.069 & LTC-BCH & -2.44 & -2.46 \\
            7     & LINK-TRX & 0.097 & 0.051 & LTC-BCH & -2.44 & -2.32 \\
            8     & ETH-LTC & 0.010 & 0.010 & ETH-BCH & -2.66 & -2.47 \\
            9     & ETH-LTC & 0.042 & 0.022 & LTC-BNB & -3.43 & -4.80 \\
            10    & ETH-BCH & 0.097 & 0.073 & LTC-BCH & -3.03 & -2.02 \\
            11    & LTC-BCH & 0.079 & 0.010 & LTC-BCH & -2.58 & -2.63 \\
            12    & ATOM-ADA & 0.010 & 0.010 & BCH-ATOM & -1.96 & -3.02 \\
            13    & BAT-ONT & 0.095 & 0.021 & ADA-XTZ & -2.61 & -2.07 \\
            14    & ADA-EOS & 0.064 & 0.010 & LTC-ADA & -2.15 & -3.23 \\
            15    & EOS-LTC & 0.044 & 0.027 & EOS-LTC & -2.33 & -2.14 \\
            16    & BAT-IOTA & 0.022 & 0.010 & TRX-BNB & -2.56 & -2.06 \\
            17    & BNB-TRX & 0.067 & 0.036 & BCH-BNB & -1.96 & -3.08 \\
            18    & BCH-LTC & 0.082 & 0.086 & TRX-IOTA & -3.08 & -2.64 \\
            19    & ETH-TRX & 0.059 & 0.023 & DASH-BCH & -2.77 & -3.42 \\
            20    & ETH-LTC & 0.078 & 0.088 & BCH-EOS & -2.07 & -2.46 \\
            21    & TRX-LTC & 0.010 & 0.010 & TRX-LTC & -2.17 & -2.05 \\
            22    & \multicolumn{1}{c}{-} & - & - & LTC-ETC & -1.99 & -2.39 \\
            23    & \multicolumn{1}{c}{-} & - & - & BNB-LINK & -2.10 & -2.06 \\
            24    & ETH-BNB & 0.072 & 0.092 & ETH-BNB & -2.05 & -2.93 \\
            25    & LTC-EOS & 0.018 & 0.018 & LTC-BNB & -1.97 & -2.32 \\
            26    & ETH-LTC & 0.085 & 0.010 & LTC-XRP & -2.50 & -3.91 \\
            27    & BNB-DASH & 0.041 & 0.043 & XRP-DASH & -2.02 & -2.48 \\
            28    & ETH-BCH & 0.060 & 0.096 & ETH-BCH & -1.95 & -2.35 \\
            29    & BNB-LTC & 0.068 & 0.015 & BNB-LTC & -2.42 & -2.99 \\
            30    & BNB-EOS & 0.044 & 0.030 & BNB-EOS & -2.22 & -3.77 \\
            31    & ETH-LINK & 0.030 & 0.015 & LTC-LINK & -2.52 & -2.72 \\
            32    & ETH-LINK & 0.058 & 0.096 & ETH-LTC & -2.46 & -1.95 \\
            33    & LTC-XRP & 0.063 & 0.032 & LTC-BNB & -2.73 & -1.93 \\
            34    & EOS-XRP & 0.049 & 0.041 & EOS-LTC & -2.36 & -1.97 \\
            35    & ETH-EOS & 0.061 & 0.041 & EOS-BNB & -2.15 & -3.96 \\
            36    & BNB-LINK & 0.090 & 0.052 & LTC-ADA & -3.91 & -2.53 \\
            37    & ETH-BNB & 0.046 & 0.063 & ETH-ETC & -2.50 & -1.93 \\
            38    & ETH-LINK & 0.013 & 0.063 & BCH-ONT & -2.49 & -2.10 \\
            39    & DASH-ONT & 0.079 & 0.010 & LTC-DASH & -2.08 & -2.20 \\
            40    & LTC-BNB & 0.091 & 0.090 & DASH-BNB & -2.18 & -2.07 \\
            41    & ETC-DASH & 0.091 & 0.035 & ETC-DASH & -1.95 & -2.52 \\
            42    & BCH-LTC & 0.017 & 0.027 & BCH-LTC & -3.15 & -2.30 \\
            43    & ETH-ETC & 0.075 & 0.010 & ETC-EOS & -3.81 & -2.44 \\
            44    & ETH-ETC & 0.045 & 0.010 & ETH-ETC & -2.66 & -3.13 \\
            45    & EOS-ETC & 0.021 & 0.010 & EOS-ETC & -4.45 & -3.32 \\
            46    & EOS-ETC & 0.027 & 0.010 & EOS-ETC & -2.46 & -3.56 \\
            47    & ETC-XRP & 0.055 & 0.069 & ETC-XRP & -3.07 & -2.75 \\
            48    & ETH-LTC & 0.089 & 0.088 & LTC-DASH & -3.29 & -4.01 \\
            49    & ETH-ETC & 0.026 & 0.010 & ETH-ETC & -2.43 & -2.18 \\
            50    & ETH-ETC & 0.070 & 0.010 & ETH-ETC & -2.96 & -2.75 \\
            51    & ETC-LTC & 0.013 & 0.020 & ETH-ETC & -2.43 & -3.19 \\
            52    & LTC-EOS & 0.010 & 0.050 & LTC-EOS & -3.39 & -2.61 \\
            \bottomrule
            \end{tabular}}%
          \label{tab 7}%
        \end{table}%
        \begin{table}[H]
            \caption{\small \textbf{Selected pairs using unit-root tests and Kendall's $\tau$ coefficient (part II)}}
          \centering
          \resizebox{0.8\textwidth}{!}{%
            \begin{tabular}{r|lrr|lrr}
            \toprule
                  & \multicolumn{3}{c|}{\textbf{ADF Unit-Root Test Result }} & \multicolumn{3}{c}{\textbf{KSS Unit-Root Test Result}} \\
            \multicolumn{1}{c|}{\textbf{Week}} & \multicolumn{1}{c}{\textbf{Pair}} & \multicolumn{1}{c}{\textbf{P-Value ($S_1$) }} & \multicolumn{1}{c|}{\textbf{P-Value ($S_2$) }} & \multicolumn{1}{c}{\textbf{Pair}} & \multicolumn{1}{c}{\textbf{t-stat ($S_1$) }} & \multicolumn{1}{c}{\textbf{t-stat ($S_2$) }} \\
            \midrule
            53    & EOS-XRP & 0.032 & 0.010 & EOS-XRP & -2.43 & -3.51 \\
            54    & \multicolumn{1}{c}{-} & - & - & LTC-BNB & -2.03 & -2.24 \\
            55    & DASH-XLM & 0.041 & 0.010 & EOS-LTC & -2.20 & -2.01 \\
            56    & ETH-BNB & 0.049 & 0.016 & ETH-LTC & -2.69 & -2.77 \\
            57    & ETH-BNB & 0.010 & 0.018 & ETH-BNB & -2.83 & -3.20 \\
            58    & ETH-BNB & 0.028 & 0.023 & ETH-BNB & -2.78 & -4.00 \\
            59    & ETH-BAT & 0.024 & 0.068 & ETH-BAT & -3.41 & -1.96 \\
            60    & ETH-BNB & 0.020 & 0.020 & ETH-EOS & -2.41 & -1.95 \\
            61    & ETH-BNB & 0.085 & 0.010 & BNB-ADA & -3.02 & -2.04 \\
            62    & BNB-ONT & 0.010 & 0.083 & ETH-BNB & -1.95 & -2.79 \\
            63    & BNB-LTC & 0.060 & 0.022 & LTC-LINK & -3.47 & -2.32 \\
            64    & LINK-ONT & 0.029 & 0.095 & XRP-LINK & -2.14 & -2.34 \\
            65    & XRP-XLM & 0.010 & 0.095 & XRP-ADA & -3.35 & -2.16 \\
            66    & ETH-BNB & 0.010 & 0.047 & ETH-BNB & -3.15 & -3.10 \\
            67    & ETH-BNB & 0.041 & 0.010 & ETH-BNB & -3.40 & -3.23 \\
            68    & ETH-BNB & 0.017 & 0.010 & ETH-BNB & -3.56 & -3.42 \\
            69    & \multicolumn{1}{c}{-} & - & - & BCH-XLM & -2.27 & -2.03 \\
            70    & BNB-ETC & 0.031 & 0.033 & LINK-BNB & -2.07 & -2.39 \\
            71    & LINK-ETC & 0.070 & 0.062 & ETC-BNB & -2.29 & -2.42 \\
            72    & LINK-EOS & 0.046 & 0.025 & BNB-LINK & -2.28 & -2.37 \\
            73    & DASH-ETC & 0.037 & 0.044 & EOS-DASH & -1.96 & -3.22 \\
            74    & ONT-ZEC & 0.026 & 0.010 & BNB-DASH & -2.56 & -1.95 \\
            75    & ZEC-BCH & 0.047 & 0.041 & ETH-BNB & -2.36 & -2.16 \\
            76    & ETH-XTZ & 0.010 & 0.029 & ETH-ZEC & -2.83 & -2.52 \\
            77    & ETH-BNB & 0.057 & 0.064 & ETH-BNB & -2.77 & -2.42 \\
            78    & ETH-BNB & 0.022 & 0.010 & ETH-BNB & -3.17 & -2.19 \\
            79    & BNB-LTC & 0.021 & 0.037 & BNB-LTC & -2.93 & -3.50 \\
            80    & BNB-LTC & 0.017 & 0.061 & BNB-LTC & -3.08 & -2.09 \\
            81    & ETH-LTC & 0.010 & 0.010 & LTC-LINK & -3.25 & -2.52 \\
            82    & LTC-XRP & 0.014 & 0.017 & LTC-XRP & -2.68 & -4.02 \\
            83    & LTC-DASH & 0.010 & 0.047 & LTC-DASH & -4.27 & -2.64 \\
            84    & ETH-LTC & 0.044 & 0.010 & ETH-LTC & -2.40 & -3.95 \\
            85    & ETH-IOTA & 0.090 & 0.010 & IOTA-DASH & -2.63 & -2.87 \\
            86    & BAT-IOTA & 0.010 & 0.010 & BAT-DASH & -3.54 & -1.96 \\
            87    & BNB-ETC & 0.023 & 0.068 & BAT-ETC & -2.49 & -2.85 \\
            88    & BNB-TRX & 0.020 & 0.084 & DASH-BAT & -2.35 & -2.43 \\
            89    & ETH-DASH & 0.090 & 0.010 & ETH-DASH & -2.71 & -4.19 \\
            90    & ETH-DASH & 0.010 & 0.014 & ETH-DASH & -2.95 & -2.78 \\
            91    & ETH-BAT & 0.095 & 0.032 & ETH-DASH & -2.54 & -2.26 \\
            92    & ETH-BAT & 0.010 & 0.084 & ETH-BAT & -3.08 & -2.15 \\
            93    & BNB-LTC & 0.014 & 0.023 & BNB-LTC & -2.22 & -3.47 \\
            94    & BCH-DASH & 0.010 & 0.046 & BCH-DASH & -3.36 & -2.30 \\
            95    & ETH-BCH & 0.051 & 0.034 & BCH-DASH & -6.40 & -2.29 \\
            96    & ETH-ADA & 0.081 & 0.010 & ETH-ADA & -2.44 & -2.53 \\
            97    & ETH-BAT & 0.042 & 0.053 & ETH-ADA & -2.47 & -2.47 \\
            98    & ADA-BAT & 0.039 & 0.010 & ETH-ADA & -1.97 & -3.08 \\
            99    & XTZ-BAT & 0.010 & 0.010 & ETH-ATOM & -1.98 & -2.15 \\
            100   & ETH-BCH & 0.010 & 0.099 & ETH-ONT & -2.20 & -3.13 \\
            101   & \multicolumn{1}{c}{-} & - & - & LINK-BCH & -2.47 & -2.00 \\
            102   & ETH-BNB & 0.057 & 0.033 & ETH-LTC & -2.48 & -2.00 \\
            103   & BNB-LINK & 0.010 & 0.020 & BNB-LINK & -3.85 & -3.27 \\
            104   & LINK-DASH & 0.061 & 0.100 & LINK-ONT & -2.49 & -2.31 \\
            \bottomrule
            \end{tabular}}%
          \label{tab 8}%
        \end{table}%
        
         \afterpage{
         \begin{landscape}
             \thispagestyle{empty}
            \begin{table}[H]
                \caption{\small \textbf{Bivariate Archimedean copulas distributions}}
                \centering
                \resizebox{1.41\textwidth}{!}{%
                \begin{tabular}{@{}llll@{}}
                    \toprule
                    \textbf{Name} & \multicolumn{1}{l}{\textbf{Bivariate Copula Distribution $\mathbf{C(u_1,u_2)\quad(\overline{u_i} = 1-u_i)}$}} & \multicolumn{1}{l}{\textbf{Generator $\mathbf{\phi(t)}$ }} & \multicolumn{1}{l}{\textbf{Parameters}}\\
                    \midrule
                    Clayton &  $\Bigg[\max \bigg(u_1^{-\theta}+u_2^{-\theta}-1,0\bigg)\Bigg]^{-1/\theta}$ &$ \frac{1}{\theta}\Big(t^{-\theta}-1\Big)$     &  $\theta >0$    \\
                    Gumbel &   $\exp{\Bigg[-\bigg[\left(-\ln{u_1}\right)^{\theta}+\left(-\ln{u_2}\right)^{\theta}\bigg]^{1/\theta}\Bigg]}$ &  $\left(-\ln(t)\right)^{\theta}$     & $\theta \geq 1$   \\
                    Frank &  $-{\theta}^{-1}\ln{\Bigg[1+\big(e^{-\theta}-1\big)^{-1}\big(e^{-\theta u_1}-1\big)\big(e^{-\theta u_2}-1\big)\Bigg]}$  &   $-\ln\Big(\frac{e^{-\theta t}-1}{e^{-\theta}-1}\Big)$    & $\theta \in \mathbbm{R}\setminus \{0\}$     \\
                    Joe   &   $1-\Bigg[\big(1-u_1\big)^{\theta}+\big(1-u_2\big)^{\theta}- \big(1-u_1\big)^{\theta}\big(1-u_2\big)^{\theta}\Bigg]^{1/\theta}$  &   $-\ln\Big(1-(1-t)^{\theta}\Big)$    &  $\theta \geq 1$ \\
                    BB1   &   $\Bigg[1+\bigg[\Big(u_1^{-\theta}-1\Big)^{\delta}+\Big(u_2^{-\theta}-1\Big)^{\delta}\bigg]^{1/\delta}\Bigg]^{-1/\theta}$  &   $\Big(t^{-\theta}-1\Big)^{\delta} $   & $\theta >0, \,\, \delta\geq 1$  \\
                    BB6   &   $1-\Bigg[1-\exp{\bigg(-\Big[\big[-\ln{\left(1-\overline{u_1}^{\,\theta}\right)}\big]^{\delta} + \big[-\ln{\left(1-\overline{u_1}^{\,\theta}\right)}\big]^{\delta} \Big]^{1/\delta}\bigg)}\Bigg]^{1/\theta}$  &   $\Big(\ln\left(1-(1-t)^{\theta}\right)\Big)^{\delta}$    & $\theta \geq 1, \,\, \delta\geq 1$   \\
                    BB7   &    $1-\Bigg[1-\bigg(\Big(1-\overline{u_1}^{\,\theta}\Big)^{-\delta}+\Big(1-\overline{u_2}^{\,\theta}\Big)^{-\delta}-1\bigg)^{-1/\delta}\Bigg]^{1/\theta}$&   $\Big(1-(1-t)^{\theta}\Big)^{-\delta}-1$    & $\theta \geq 1, \,\, \delta>0$\\
                    BB8   &  $\delta^{-1}\Bigg[1-\left(1-\left[1-\left(1-\delta\right)^{\theta}\right]^{-1}\left[1-\left(1-\delta u_1\right)^{\theta}\right]\left[1-\left(1-\delta u_2\right)^{\theta}\right]\right)^{1/\theta}\Bigg]$ &   $-\ln\bigg(\frac{1-(1-\delta t)^{\theta}}{1-(1-\delta)^{\theta}}\bigg)$    & $\theta \geq 1, \,\,0 < \delta \leq 1$\\
                    \bottomrule
                \end{tabular}}%
                \label{tab 3}%
            \end{table}%
         \end{landscape}
         }
\end{document}